\documentclass[draft]{agujournal2019}
\usepackage{apacite}
\usepackage{url} %this package should fix any errors with URLs in refs.
\usepackage{lineno}
\usepackage{aas_macros}
\usepackage{siunitx}
\usepackage{comment}

\newcommand{\Bz}{B$_\textrm{z}$}
\newcommand{\Bt}{B$_\textrm{t}$}

\draftfalse
\journalname{Space Weather}

\begin{document}

\title{Machine learning for predicting the B$_\textrm{z}$ magnetic field component from upstream in situ observations of solar coronal mass ejections}

\authors{M.~A.~Reiss\affil{1}, C.~M\"ostl\affil{1}, R.~L.~Bailey\affil{2}, H.~T.~R\"udisser\affil{3,4}, U.~V.~Amerstorfer\affil{1}, T.~Amerstorfer\affil{1}, A.~J.~Weiss\affil{1,4}, J.~Hinterreiter\affil{1,4}, and A.~Windisch\affil{3,5,6}}

\affiliation{1}{Space Research Institute, Austrian Academy of Sciences, Schmiedlstraße 6, 8042 Graz, Austria}
\affiliation{2}{Zentralanstalt f\"ur Meteorologie und Geodynamik, Hohe Warte 38, 1190 Vienna, Austria}
\affiliation{3}{Know-Center GmbH, Inffeldgasse 13, 8010 Graz, Austria}
\affiliation{4}{Institute of Physics, University of Graz, Universit\"atsplatz 5, 8010 Graz, Austria}
\affiliation{5}{Department of Physics, Washington University in St. Louis, MO 63130, USA}
\affiliation{6}{RL Community, AI AUSTRIA, Wollzeile 24/12, 1010 Vienna, Austria}

\correspondingauthor{Martin A. Reiss}{martin.reiss@oeaw.ac.at}

\begin{keypoints}
\item We hypothesize that upstream in situ measurements are sufficient to predict {the minimum value of the B$_\textrm{z}$ component} in solar coronal mass ejections.
\item We present a predictive tool that forecasts the minimum of B$_\textrm{z}$ in an ICME with a MAE of $3.12$~nT and a PCC of $0.71$.
\end{keypoints}

\begin{abstract}
Predicting the B$_\textrm{z}$ magnetic field embedded within ICMEs, also known as the B$_\textrm{z}$ problem, is a key challenge in space weather forecasting. We study the hypothesis that upstream in situ measurements of the sheath region and the first few hours of the magnetic obstacle provide sufficient information for predicting the downstream \Bz{} component. 

To do so, we develop a predictive tool based on machine learning that is trained and tested on 348~ICMEs from Wind, STEREO-A, and STEREO-B measurements. We train the machine learning models to predict the minimum value of the B$_\textrm{z}$ component and the maximum value of the total magnetic field B$_\textrm{t}$ in the magnetic obstacle. To validate the tool, we let the ICMEs sweep over the spacecraft and assess how continually feeding in situ measurements into the tool improves the B$_\textrm{z}$ prediction. 

{We specifically find that} the predictive tool can predict the minimum value of the B$_\textrm{z}$ component in the magnetic obstacle with a mean absolute error of $3.12$~nT and a Pearson correlation coefficient of $0.71$ when the sheath region and the first 4~hours of the magnetic obstacle are observed. While the underlying hypothesis is unlikely to solve the B$_\textrm{z}$ problem, the tool shows promise for ICMEs that have a recognizable magnetic flux rope signature. Transitioning the tool to operations could lead to improved space weather forecasting.
\end{abstract}

\section*{Plain Language Summary}
At any time, our solar system is populated with interplanetary coronal mass ejections (ICMEs). Solar scientists and space weather forecasters track ICMEs when they are ejected from the Sun and follow their path into the vast reaches of interplanetary space. They do so because if an ICME hits Earth, it could damage our infrastructure such as power-grids and GPS satellites, which are a mainstay of our modern civilization. The possible damage is primarily determined by the magnetic field embedded within the ICME. The North-South magnetic field component, B$_\textrm{z}$, plays a decisive role, especially if it is pointing opposite to Earth's magnetic field. Currently we cannot predict B$_\textrm{z}$ with sufficient accuracy. Scientists often refer to our limited predictive abilities as the B$_\textrm{z}$ problem. Here we shine a new light on the B$_\textrm{z}$ problem by developing a predictive tool based on machine learning that is trained and tested on 348~ICMEs. By feeding measurements of the ICME into machine learning algorithms, we find that our predictive tool can forecast the B$_\textrm{z}$ component reasonably well. While our tool does not solve the B$_\textrm{z}$ problem, it shows promise in forecasting and potentially mitigating the effects of ICMEs on our planet Earth and its inhabitants.

\section{Introduction}
The B$_\textrm{z}$ component of the interplanetary magnetic field (IMF) largely determines the amount of energy and momentum transferred from the solar wind into the Earth's magnetosphere via magnetic reconnection at the dayside magnetopause~\cite{dungey1961}. {Knowledge of the future B$_\textrm{z}$ magnitude} is integral for monitoring and predicting the energy input into the magnetosphere, ionosphere, and thermosphere. There is a current lack of B$_\textrm{z}$ forecasting capabilities, which is often referred to as the B$_\textrm{z}$ problem. In essence, the field of space weather forecasting needs innovation to predict the north-south component (B$_\textrm{z}$) of the IMF in near-Earth space during intense geoeffective events.

Key criteria for intense geoeffective events are extended periods of large southward B$_\textrm{z}$, which points opposite to the Earth's magnetic field~\cite{gonzalez1987}. The largest southward B$_\textrm{z}$ disturbances in the IMF are found in coronal mass ejections (CMEs), plasma clouds with embedded magnetic fields that are expelled from the Sun into our solar system~\cite<see>[for a review]{Webb_2012}. At any time, the heliosphere is populated with interplanetary coronal mass ejections (ICMEs) that interact with the prevailing ambient solar wind flows and fields. Most of the structure in the ambient solar wind comes from interacting fast and slow solar wind flows~\cite{owens13}. Fluctuating magnetic fields in so-called stream interaction regions can also cause weak to moderate geomagnetic activity~\cite{zhang2007, kilpua2017a}. Although stream interaction regions are a driver of recurrent geomagnetic activity~\cite{tsurutani06}, the most extreme geomagnetic disturbances are caused by large-scale B$_\textrm{z}$ perturbations embedded within ICMEs~\cite{echer2008}. It is primarily during ICME events, when accurate B$_\textrm{z}$ estimates are needed most, that the prediction thereof is most difficult and the B$_\textrm{z}$ problem is most relevant.

Breakthroughs in the B$_\textrm{z}$ problem are challenging for many reasons, including observational limitations~\cite<see>{vourlidas2019}. First, we can not accurately deduce the magnetic properties of CMEs such as magnetic field magnitude, topology, and helicity at the time of formation from observations alone. Radio and EUV off-limb imaging and spectroscopy can inform physics-based models, but we cannot measure these magnetic properties directly. Second, the coronal conditions within the Alfv\'en surface below approximately 20~solar radii, through which the CME evolves, are unknown. Due to the unknown condition in the corona, we do not understand the origin and early evolution of CMEs well enough to predict their magnetic structure~\cite{vourlidas2013}. Third, we cannot monitor the rotation, compression, deflection, and reconnection of ICMEs in interplanetary space with ambient solar wind fields, which influence the ICME magnetic field structure before they arrive at Earth. 

When observations are limited, numerical models can provide insights into the physical conditions. One promising research avenue to tackle the B$_\textrm{z}$ problem focuses on solving the three-dimensional equation of magnetohydrodynamics (MHD). Examples for MHD codes that incorporate the CME magnetic field structure are SUSANOO-CME~\cite{shiota2016}, AWSoM-SWMF~\cite{jin2017}, MAS~\cite{torok2018} and EUHFORIA \cite{poedts_2020}. Although a full physical description is desirable, solving the B$_\textrm{z}$ problem with MHD codes is challenging, as will be discussed later.

Without a definitive physical solution, we need to study new predictive tools that can enhance our predictive capabilities, and, ideally, inform the boundary conditions of full physics-based models. Today's predictions of the B$_\textrm{z}$ component at Earth rely on empirical relationships, statistical extrapolation, pattern recognition, machine learning algorithms, and more. \citeA{chen96} and \citeA{chen1997} first introduced the idea to use the coherence of magnetic flux ropes in ICMEs to predict their magnetic structure. They found that the B$_\textrm{z}$ component at Earth can be predicted with lead times of up to 10~hours for well-defined magnetic clouds with a smooth field rotation. \citeA{kim2014} focused on early CME properties and updated their prediction by monitoring the solar wind conditions in space. In contrast, \citeA{savani2015} used remote sensing for predicting the magnetic field components in magnetic clouds at Earth. Here, their prediction relies on the categories introduced in~\citeA{bothmer1998} and \citeA{mulligan1998}, among other things. More recently, \citeA{riley2017} developed a pattern recognition technique for predicting B$_\textrm{z}$, and \citeA{owens2017} used a past analogs method to predict the conditions in the ambient solar wind flow and geomagnetic indices at Earth. \citeA{salman18} studied historic events to predict the southward interplanetary B$_\textrm{z}$ periods after interplanetary shocks. \citeA{Moestl_2018} used solar observations to determine the initial state of the CME, assuming a self-similar expansion. In contrast, \citeA{dossantos2020} developed a deep neural network and an analytical flux rope model to identify the internal structure of ICMEs. 

To date, however, scant attention has been paid to using upstream in situ measurements for predicting the B$_\textrm{z}$ of the magnetic obstacle embedded within ICMEs at the Sun-Earth L1 point. In this study, we present a predictive tool based on machine learning {that simplifies the} B$_\textrm{z}$ problem by predicting estimates of the B$_\textrm{z}$ component for the whole magnetic flux rope, particularly min(B$_\textrm{z}$) and max(B$_\textrm{t}$), which can be used as an estimate of the maximum B$_\textrm{z}$ component. We train, test, and validate the predictive tool on bulk plasma and in situ measurements from 348 ICMEs observed close to a heliocentric distance of 1 AU~\cite{Moestl2017, Moestl2020}. 

To simulate an operational space weather forecast, we feed in situ data into the predictive tool as if the ICMEs sweep over the spacecraft in an experimental real-time mode. We then assess how progressively adding data from the ICME sheath and parts of the magnetic obstacle improves the predictive skill. {While the predictive tool is far from solving} the B$_\textrm{z}$ problem, it shows reasonable first results in predicting estimates for the lower limit of the B$_\textrm{z}$ component in ICMEs. 

For the sake of consistency and clarity in this study, we follow the definition in~\citeA{nieves_chinchilla_2018} and use the term 'magnetic obstacle' to refer to the magnetic structure embedded in an ICME, which can deviate from in situ signatures of an idealized magnetic flux rope in~\citeA{Burlaga_1981}. We furthermore refer to the sheath as the region of compressed solar wind between the ICME shock front and the leading edge of the magnetic obstacle~\cite{owens05} and use the term ICME for the interval of disturbed solar wind conditions including the sheath region and the magnetic obstacle~\cite{rouillard11}. 

The paper consists of the following parts. After discussion of the data sources in Section~\ref{sec:data}, Section~\ref{sec:machinelearning} outlines the machine learning approach, Section~\ref{sec:validation} outlines the {validation analysis, and} Section~\ref{sec:results} presents the skill of the predictive tool. Section~\ref{sec:discussion} discusses our findings and outlines future perspectives, and Section~\ref{sec:summary} summarizes the study. Section~\ref{sec:sources} links to our publicly available online resources and the ICME catalog, in situ solar wind data, and the source code. All the machine learning algorithms are taken from the Python packages Scikit-learn and Keras, which are open-source, straightforward to use, and thoroughly tested.

\section{Data}\label{sec:data}
At any point in time, the heliosphere is populated by ICMEs, and their physical properties are continually recorded in ICME catalogs. {In this study, we use the ICMECAT, an ICME catalog that was originally created during the HELCATS project~\cite{Moestl2017}. We have published a major update in \citeA{Moestl2020} and now call it the HELIO4CAST ICME catalog or ICMECATv2.0. Here, we use the version last updated on 2021 April 29. It includes 558~ICMEs observed close to 1 AU, at the Sun-Earth L1 point and by STEREO-A/B, during the time interval 2007 January~1 to 2021 April~1. ICME events are added manually to this living catalog, guided by the criteria in \citeA{nieves_chinchilla_2018}. It contains only events that show clear signatures of magnetic structure, called magnetic obstacles. A large number of events, frequent updates, and open access (see Section~\ref{sec:sources}) are the main advantages and the reasons for this choice. }

\begin{figure}
\includegraphics[width=0.99\textwidth]{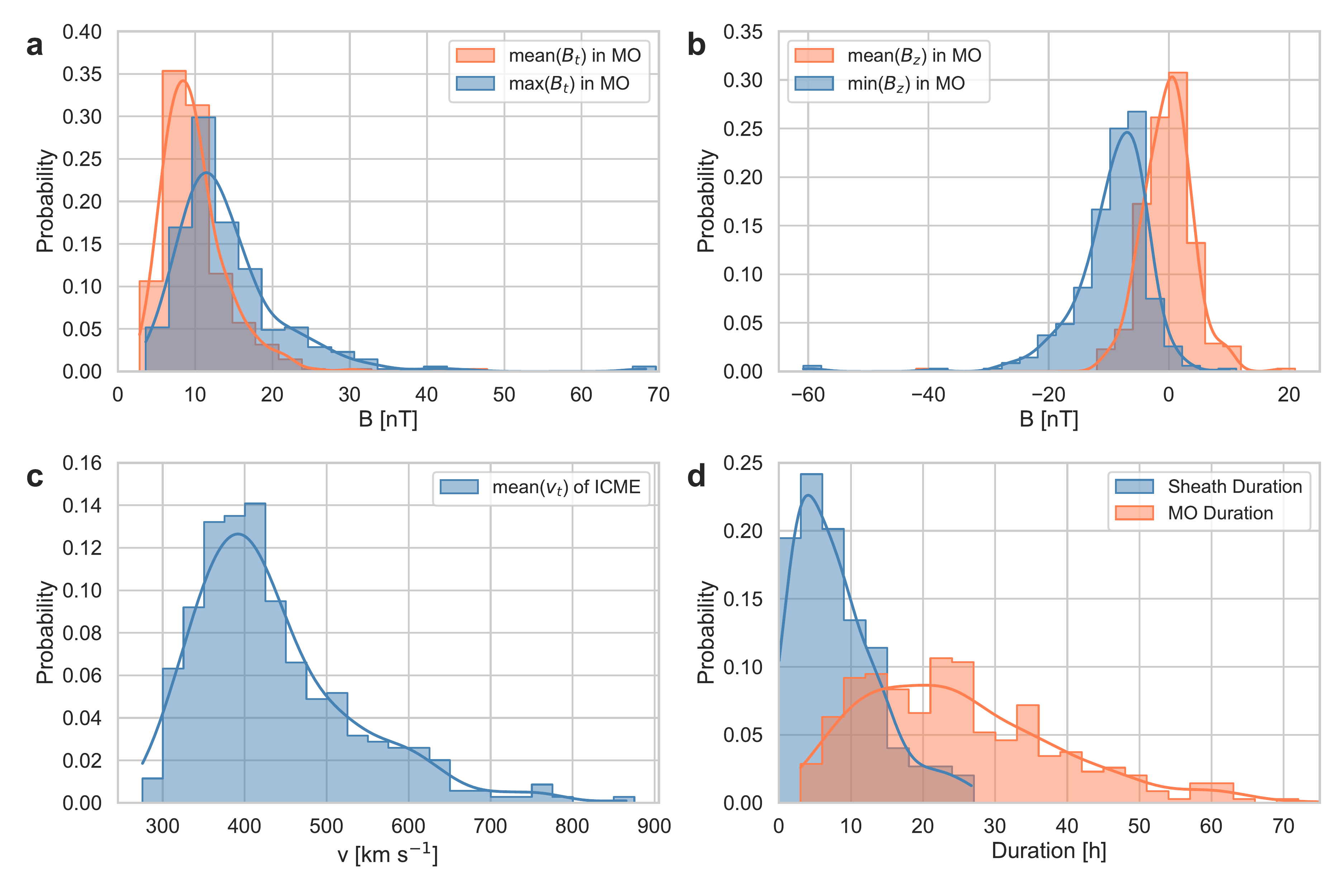}
\caption{The probability distribution functions of the physical properties of 348~ICMEs under scrutiny. (a) The mean and maximum total magnetic field in the magnetic obstacle; (b) the minimum and mean B$_\textrm{z}$ in the magnetic obstacle; (c) the mean bulk plasma speed in the ICME including the sheath and magnetic obstacle; and (d) the sheath and magnetic obstacle durations.}
\label{fig:dist}
\end{figure}

{To train, validate, and test machine learning algorithms for predicting B$_\textrm{z}$, we use bulk plasma and in situ measurements for ICMEs in ICMECATv2.0 that are close to a heliocentric distance of 1~AU. In particular, we study in situ measurements from the MFI and SWE instruments on the Wind spacecraft~\cite{ogilvie1995, lepping1995} and the IMPACT and PLASTIC instruments on the STEREO-A and STEREO-B spacecraft~\cite{Luhmann_2008,galvin2008}. From the 558~ICMEs in ICMECATv2.0 observed by WIND, STEREO-A, and STEREO-B, we focus on 362~ICMEs (or $65\%$) that show either sheath region signatures or a density pileup in front of a magnetic obstacle. This selection results in 149~ICMEs in Wind, 135~ICMEs in STEREO-A, and 78~ICMEs in STEREO-B. After cleaning the data and removing events with too many missing data points in the measurements, we end up with 348~ICMEs, with 149~ICMEs in Wind, 123~ICMEs in STEREO-A, and 76~ICMEs in STEREO-B.}

{Figure~\ref{fig:dist} shows the probability distribution functions of physical properties of the 348~ICMEs. These physical properties include the mean and maximum of the total magnetic field (B$_\textrm{t}$) in the magnetic obstacle, mean and minimum of B$_\textrm{z}$ in the magnetic obstacle, mean of the bulk plasma speed ($v_t$) in the ICME, and the duration of the sheath and magnetic obstacle.} The value range of ICME and magnetic obstacle properties that we use for machine learning is as follows. The average duration of the sheath region is $9.0 \pm 6.0$~h, and that of the magnetic obstacle is $25.2 \pm 14.6$~h. {The sheath region is defined as the interval between the \textit{icme\_start\_time} and the \textit{mo\_start\_time} parameters in the ICME catalog, and the magnetic obstacle ranges from \textit{mo\_start\_time} to \textit{mo\_end\_time}.} The mean maximum value of B$_\textrm{t}$ in the magnetic obstacle is $14.3 \pm 7.6$~nT, and the mean minimum value of B$_\textrm{z}$ in the magnetic obstacle is $-9.4 \pm 6.8$~nT.

\begin{figure}
\includegraphics[width=0.99\textwidth]{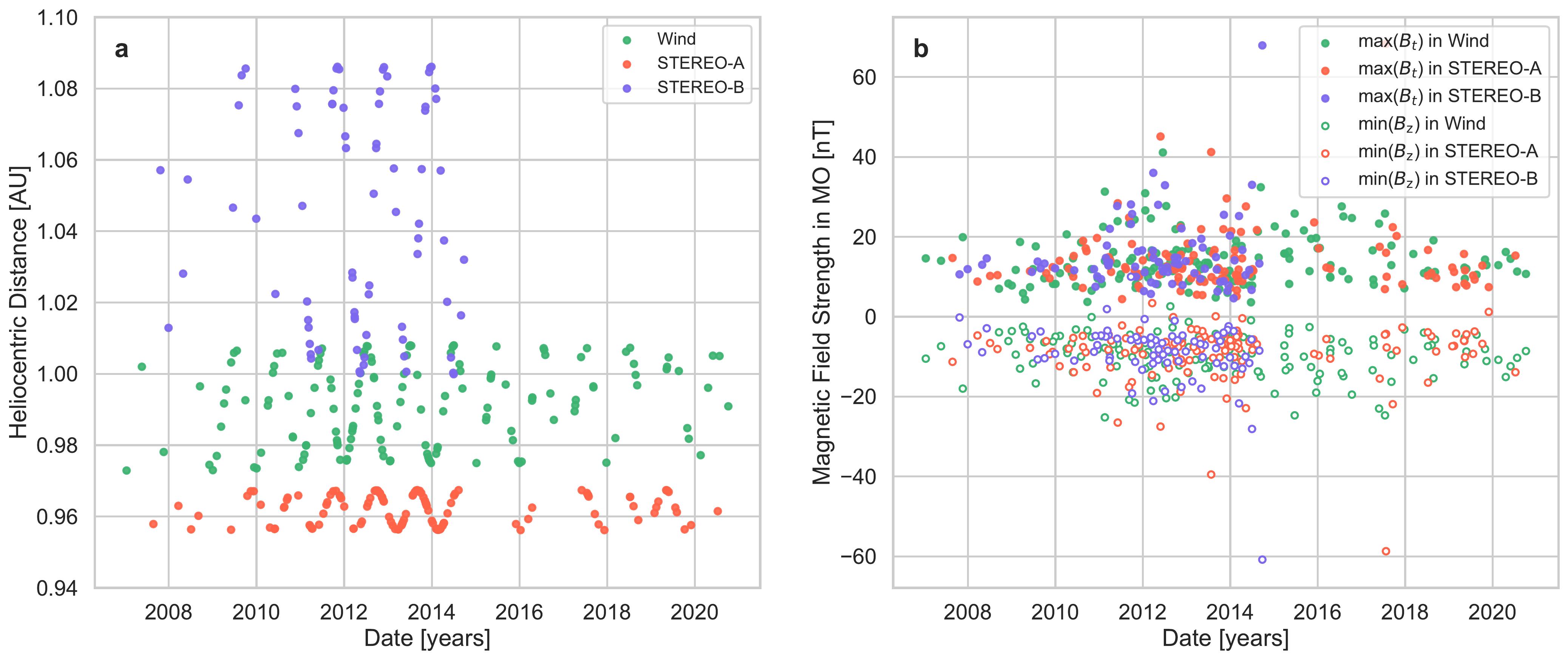}
\caption{Properties of the 348~ICMEs in the catalog. (a) Heliocentric distance for each ICME event as a function of time, observed with Wind (green), STEREO-A (red), and STEREO-B (blue); (b) Maximum total magnetic field (filled circles) B$_\textrm{t}$ and minimum value of the B$_\textrm{z}$ component (open circles) in the magnetic obstacle (MO) of the ICME for each event as a function of time.}
\label{fig:icmecat}
\end{figure}

{Figure~\ref{fig:icmecat}a} shows the heliocentric distance as a function of time for the 348 ICMEs, with the color indicating the observing spacecraft. {Figure~\ref{fig:icmecat}b} illustrates the maximum values of the total magnetic field B$_t$ and the minimum of the B$_\textrm{z}$ component in the magnetic obstacles. These two properties are the targets we want to predict. {To do so, we train machine learning algorithms with ICME properties including B$_\textrm{t}$, $v_t$, the magnetic field components (B$_\textrm{x}$, B$_\textrm{y}$, B$_\textrm{z}$), and the proton temperature ($T_p$) and density ($N_p$).} 

%Statistics for the final 348 selected events with sheath:
%Average ICME length   : 34.26 hours
%Average MO length     : 25.24 hours
%Average SHEATH length : 9.02 hours

%STD ICME length       : 17.02 hours
%STD MO length         : 14.60 hours
%STD SHEATH length     : 6.02 hours

%Average MO Bt max   : 14.32 nT
%std MO Bt max   : 7.57 nT

%Average MO Bt   : 9.95 nT
%std MO Bt   : 4.47 nT

%Average MO Bz   : -0.32 nT
%std MO Bz   : 4.65 nT

%Average MO Bzmin   : -9.43 nT
%std MO Bzmin   : 6.79 nT

\section{Machine Learning}\label{sec:machinelearning}
\begin{figure}
\includegraphics[width=0.99\textwidth]{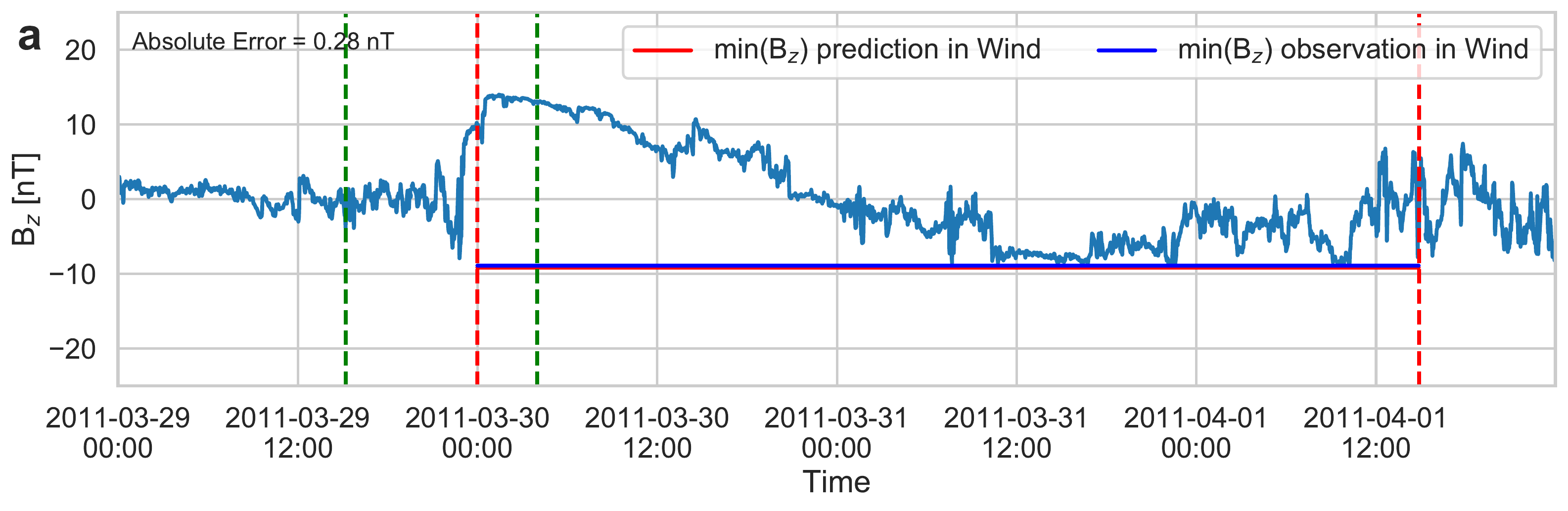}
\includegraphics[width=0.99\textwidth]{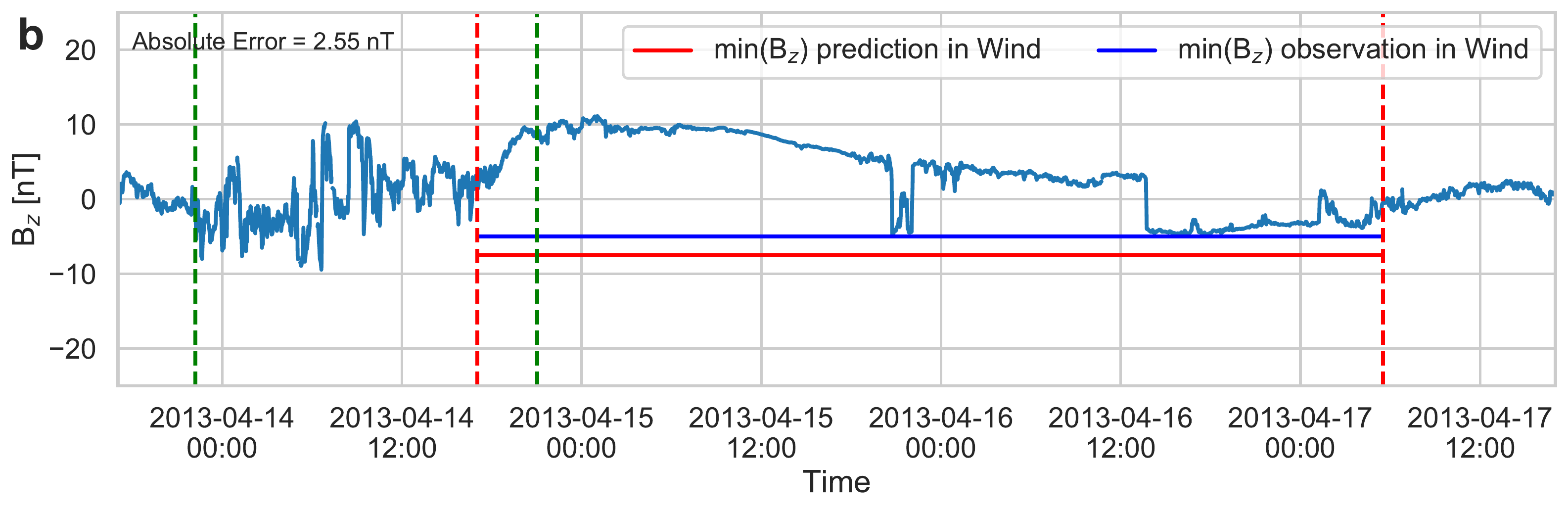}
\includegraphics[width=0.99\textwidth]{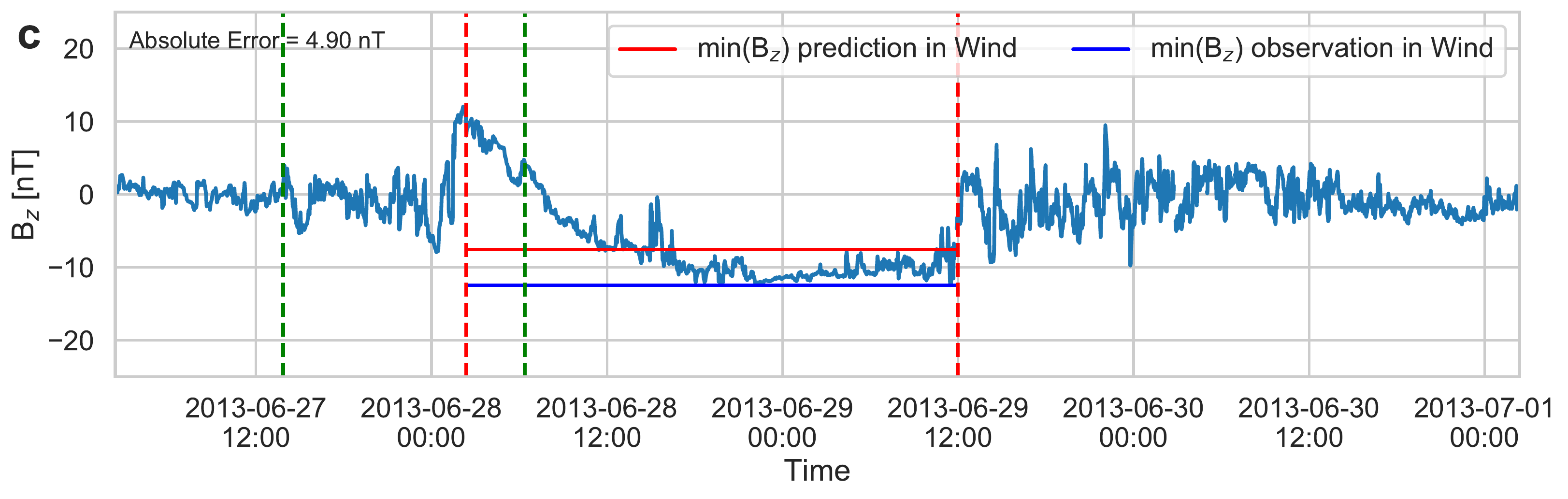}
\caption{Comparison of the predicted minimum value of the B$_\textrm{z}$ component in the magnetic obstacle using a Gradient Boosting Regressor (red) with the observation (blue). The absolute errors are close to the ideal prediction (a), median validation error (b), and the 75\%~percentile in the absolute error (c), respectively. The {vertical} green dashed lines mark the time interval where we compute the feature values, and the {vertical} red dashed lines mark the magnetic obstacle where we predict B$_\textrm{z}$. {In these three examples, we use the ICME sheath and the first 4~hours of the magnetic obstacle indicated by the green dashed lines as input for the min(B$_\textrm{z}$) prediction indicated by the horizontal red line.}}
\label{fig:mfrPred}
\end{figure}

To predict the B$_\textrm{z}$ component in the magnetic obstacle embedded within an ICME, we train machine learning algorithms on the properties of 348~ICMEs. {Figure~\ref{fig:mfrPred}} illustrates the main idea. Feature values computed in the ICME sheath region and parts of the magnetic flux rope (window delineated by green vertical lines) are used as input to machine learning to predict the minimum of the B$_\textrm{z}$ component in the magnetic obstacle (red horizontal line). The three events in {Figure~\ref{fig:mfrPred}(a--c)} represent ideal, average, and poor B$_\textrm{z}$ predictions of the final tool, where the red horizontal lines show the prediction and the blue horizontal lines show the observation. The ideal, poor, and average examples correspond to an absolute error close to zero, the median absolute error, and the 75\%~percentile in the validation. 

In the following, we introduce the predictive tool. To do so, we specify the features and targets, study different machine learning algorithms, split the ICME catalog into training and testing sets, perform hyperparameter tuning {to optimize the machine learning algorithms}, and test the B$_\textrm{z}$ prediction in an experimental real-time mode.

\subsection{Features and Targets}\label{sec:features}
We compute the features for machine learning from in situ plasma and magnetic field measurements of 348~ICMEs. {In particular, we examine the timelines of 7 different physical properties (B$_\textrm{t}$, B$_\textrm{x}$, B$_\textrm{y}$, B$_\textrm{z}$, $v_t$, $T_p$, $N_p$) and compute 6 statistical measures for each of them}. The measures that serve as features for machine learning include the mean value, standard deviation, minimum and maximum values, the ratio between the maximum and minimum values, and the ratio between the mean value and standard deviation, also known as the coefficient of variation. These features are calculated for either the sheath region alone or the sheath region plus several hours of the magnetic obstacle interval (for example, using the first 4~hours into the magnetic obstacle, as illustrated before). {We moreover studied combinations of the physical properties, including $v_t \textrm{B}_\textrm{z}$, $v_t \textrm{B}_\textrm{t}$, $v_t^2$, B$_\textrm{t}^2$, but did not see any significant improvement in the validation metrics.}

By computing feature values for all selected ICME events, we create a $42\times348$ feature matrix with $14616$ entries {that serves as input to the machine learning algorithms.} We define two targets that we want to predict inside the magnetic obstacle: (1) the minimum of the B$_\textrm{z}$ component, min(B$_\textrm{z}$); and (2) the maximum of the total magnetic field, max(B$_\textrm{t}$).

\subsection{Machine Learning Algorithms}\label{sec:mlalgorithms}
We use a linear regressor (LR), random forest regressor (RFR), and gradient boosting regressor~\cite<GBR, see>{Friedman2001} from the widely-applied Python package Scikit-Learn to train a model. We select the RFR and the GBR from many learners because they provide the most accurate predictions for the problem at hand, are robust algorithms for training, and are efficient to implement and tune. {The RFR and GBR both rely on decision trees, which show powerful predictive skill when combined in an ensemble of hundreds of trees, usually referred to as a forest.} We also provide the results from a {simple LR that acts as a benchmark} against which future versions of the predictive tool and studies can be easily compared.

\subsection{Training and Testing}
{For every ICME, we have computed 6~statistical measures for 7~physical properties, resulting in a total of 42~features. We identify those features that affect the prediction the most. To validate the real-world performance, we split the input data into training and testing data. This final hold-out testing set is not used in training. By using the testing set for assessing the prediction, we attempt to extrapolate the model's skill to unseen behavior. The training set includes 243~ICMEs, and the testing set includes 105~ICMEs, randomly selected. This selection corresponds to 70\% and 30\% of all events. During training, we apply stratified 5-fold cross validation, in which a randomly selected subset (1/5th) of the training data set is kept aside while the model is trained on the remaining 4/5ths. Each `fold' of the data is cycled through to produce five trained models. Out of the five trained models, the one that performs best on the unseen data from the reserved subset is kept as the final model. We also apply early stopping, where model training is stopped some steps before it reaches the minimum in the loss function to prevent over-fitting to the data.}

\subsection{Hyperparameter Tuning}
We perform hyperparameter tuning for the RFR and GBR for the selected set of features. To do so, we use a grid search and test combinations of parameter sets. {The parameters for the two models include the learning rate, the maximum depth of each tree, minimum number of samples to define a node split, minimum number of samples per leaf, and the number of trees/estimators, resulting in 4 parameters for GBR tuning and 5 parameters for RFR tuning. We examine every point in this 4D (5D) parameter space within specified parameter ranges and use each set of parameters to train the GBR (RFR). The parameter set with the best scoring measure is used for training.} After the grid search, we create a model including 200~decision trees (300), where the maximum depth of each tree is 3 (5) for the GBR (RFR). The minimum number of samples to define a node split or set a leaf was also defined in hyperparameter tuning for the RFR and GBR.

\subsection{Experimental Real-Time Mode}

{We furthermore attempt to simulate an operational space weather forecast. To do so, we feed the features from the magnetic sheath and magnetic obstacle (Section~\ref{sec:features}) into three machine learning algorithms (Section~\ref{sec:mlalgorithms}) to predict max(B$_\textrm{t}$) and min(B$_\textrm{z}$) as if an ICME sweeps over the spacecraft in an experimental real-time mode. We start with the features computed from the magnetic sheath and progressively add in situ measurements from the magnetic obstacle in a 1~hour cadence. This approach results in a string of separately trained algorithms applied in succession for each input time window. Conducting such an analysis is useful because it allows us to estimate the reliability of the predictive tool in an operational situation, and furthermore points to possible weaknesses in the methodology.}

{In this way, we study how progressively adding in situ measurements from the magnetic obstacle improves the B$_\textrm{z}$ prediction. We assess how many hours from the magnetic sheath and magnetic obstacle are required to achieve a certain accuracy in magnetic obstacle max(B$_\textrm{t}$) and min(B$_\textrm{z}$) predictions.} 

\begin{table}[]
\caption{Overview of point-to-point comparison metrics.}
\begin{center}
\begin{tabular}{lll}
\hline
Metric                 & Short Name & Definition \\ \hline
Mean error             & ME         & $\frac{1}{n} \sum_{k=1}^n (f_k - o_k)$    \\ \\
Mean square error      & MSE        & $\frac{1}{n} \sum_{k=1}^n (f_k - o_k)^2$           \\ \\
Mean absolute error    & MAE        & $\frac{1}{n} \sum_{k=1}^n \left|f_k - o_k\right|$           \\ \\
Root mean square error & RMSE       & $\sqrt{\frac{1}{n} \sum_{k=1}^n (f_k - o_k)^2}$           \\ \\
Skill score            & SS         & $1 - \frac{\rm{MSE}_{\rm{pred}}}{\rm{MSE}_{\rm{ref}}}$           \\ \hline
\end{tabular}%
\end{center}
\label{tab:errorfunctions}
\end{table}

% ----------------------------------------------------------------------
\section{Validation Analysis and Metrics}\label{sec:validation}
% ----------------------------------------------------------------------
We assess the skill of the machine learning algorithms for predicting min(B$_\textrm{z}$) and max(B$_\textrm{t}$) with widely-applied validation metrics. These metrics are computed from a comparison between measurements and predictions in terms of continuous and binary variables. Contrary to continuous variables that can take on any real numbers, binary variables are categorical such as event and non-event predictions~\cite<see, for instance,>{owens05, reiss16, wold18}. Focusing on both approaches, we compute average errors from the comparison of the predicted and observed magnetic fields, and then investigate an event-based validation analysis. 

\subsection{Point-to-point Metrics}
First, we compare the predictions with observations in terms of statistical measures such as the mean, median, and standard deviation. These basic measures contain important information on the underlying statistical distribution of the predicted and observed values indicating, for example, if a model tends to over- or under-estimate the measurement. 

In addition, we study the model accuracy in terms of error functions such as the mean error (ME), mean absolute error (MAE), and the root mean square error (RMSE). Table~\ref{tab:errorfunctions} summarizes these error functions, where $(f_k,o_k)$ is the $k$-th element of $n$ pairs of forecasts and observations. Although strictly speaking not an error function, we also include the Pearson correlation coefficient (PCC) in the analysis.

Next, we examine if our predictive tool is more useful than a simple baseline model. Baseline models are a valuable diagnostic to determine the skill of a novel model approach relative to a naive prediction. Baseline models in the space weather community often rely on the mean value of past observations~\cite<see, for example,>{owens18}. For the sake of consistency, we define our baseline model as the mean value of all the targets we train our machine learning algorithms on. 

In this context, the skill score (SS) is a measure that quantifies the skill of a forecast in comparison to the baseline model. Table~\ref{tab:errorfunctions} shows the definition of SS, where $\textrm{MSE}_{\textrm{pred}}$ is the mean square error of the prediction, and $\textrm{MSE}_{\textrm{ref}}$ is the MSE of the reference baseline model. A negative SS means the model is worse than the baseline model, a SS of 0 means the model is equal to the baseline model, whereas 1 indicates an ideal prediction.

\subsection{Binary Metrics}
By categorizing each min(B$_\textrm{z}$) and max(B$_\textrm{t}$) prediction in terms of event/non-event predictions, we complement the analysis in the previous section. The advantages are plentiful as outlined in~\citeA{owens18}. First, error functions give equal importance to weak and strong magnetic field strength. Some users of our predictive tool, however, are only interested in the ability of forecasting events above a certain threshold, while smaller events are less important. Second, outliers in the prediction significantly affect point-to-point comparison metrics. For users wanting to react when the event exceeds a defined threshold, an alternative is to consider each prediction as an event/non-event prediction.

To define events and non-events in the observation and prediction, we use a {threshold value equal to the mean value of all 105~ICMEs in the test data.} Through the cross-check of event and non-event combinations in the observed and predicted MFR properties, we count the number of hits (true positives; TPs), false alarms (false positives; FPs), misses (false negatives; FNs), and correct rejections (true negatives; TNs). Summarized in the so-called contingency table, a TP is a correctly predicted event, while an FN is an event that was not predicted. On the other hand, an FP is a predicted event that was not observed, and a TN is a correctly predicted non-event. 

Table~\ref{tab:binarymetrics} shows skill measures that we compute from the entries of the contingency table. Here the ratio between the number of predictions and observations, also known as Bias (BS), shows the tendency of the machine learning algorithms to over- or under-estimate the number of events. In addition, the true skill statistics (TSS) lies in the range $[-1,1]$, where an ideal prediction would have the value 1 (or -1 for an optimum inverse prediction), and a TSS of 0 means no skill. One benefit of the TSS is that it uses all the elements in the contingency table and that it is unbiased by the proportion of predicted and observed events~\cite<see>{Hanssen1965, Bloomfield2012}. 

\begin{table}[]
\caption{Overview of binary metrics defined by the entries of a contingency table.}
\begin{center}
\begin{tabular}{lll}
\hline
Metric                 & Short Name & Definition \\ \hline
True Positive Rate     & TPR        & $\frac{\rm{TP}}{\rm{TP} + \rm{FN}}$    \\ \\
False Positive Rate    & FPR        & $\frac{\rm{FP}}{\rm{FP} + \rm{TN}}$           \\ \\
Threat Score           & TS         & $\frac{\rm{TP}}{\rm{TP} + \rm{FP} + \rm{FN}}$           \\ \\
True Skill Statistics  & TSS        & $\rm{TPR} - \rm{FPR}$           \\ \\
Bias                   & BS         & $\frac{\rm{TP} + \rm{FP}}{\rm{TP} + \rm{FN}}$           \\ \hline
\end{tabular}%
\end{center}
\label{tab:binarymetrics}
\end{table}

% ----------------------------------------------------------------------
\section{Results}\label{sec:results}
% ----------------------------------------------------------------------
We validate the predictive tool on data unused in our machine learning investigation, which is the test data set with 105~ICMEs. {To provide a fair assessment throughout the solar cycle, the 105~ICMEs are randomly selected from 2007 January 1 to 2021 April 1.} We then let each ICMEs sweep over the spacecraft in an experimental real-time mode and assess how additional information from several hours inside the magnetic flux rope improves the predictive skill.

\begin{figure}
\includegraphics[width=0.99\textwidth]{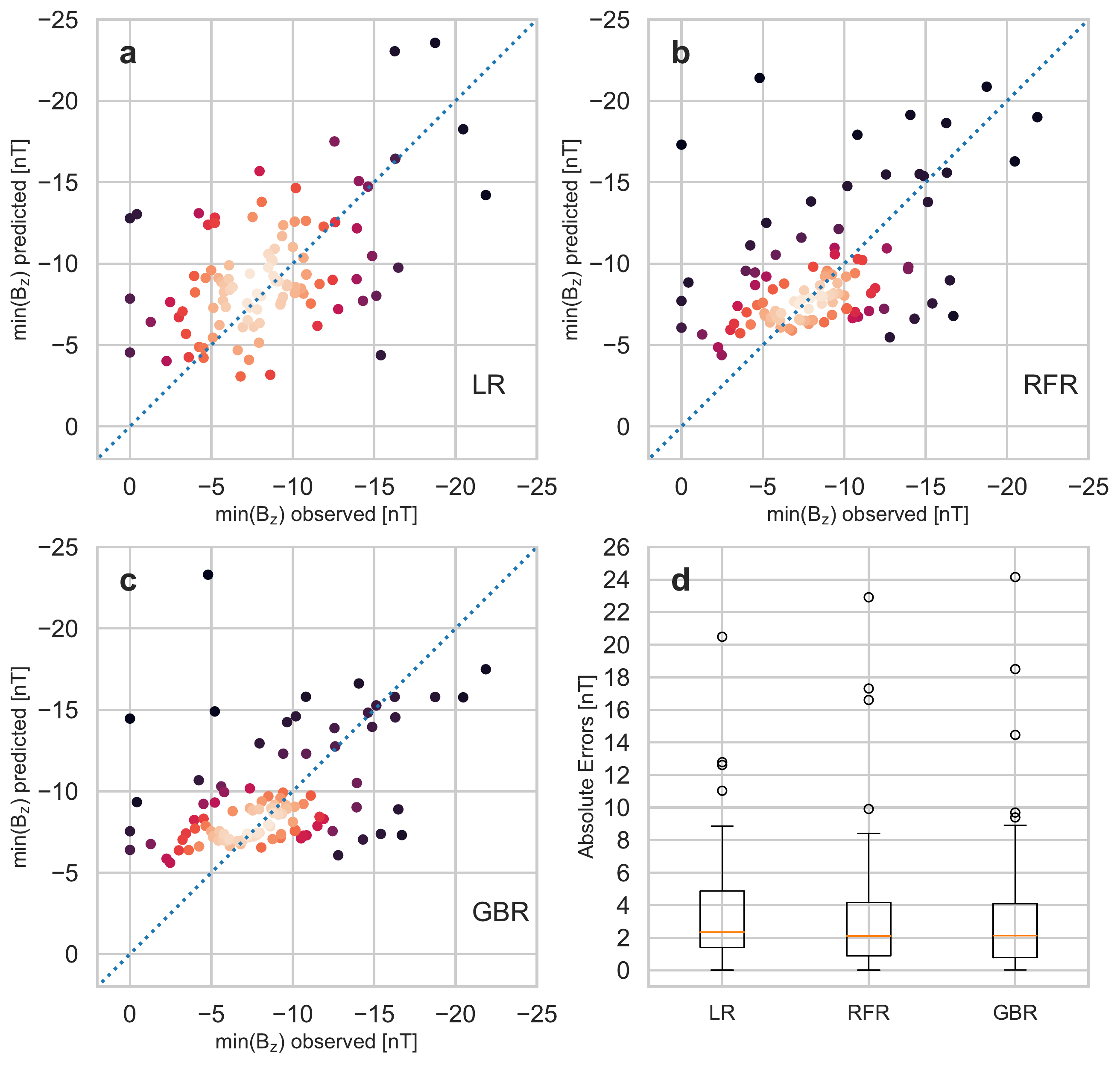}
\caption{Comparison of the predicted and observed minimum value of the B$_\textrm{z}$ component with three different machine learning algorithms, where the color scales with the density of the data points. ({a}) Linear Regressor; ({b}) Random Forest Regressor; ({c}) Gradient Boosting Regressor; ({d}) absolute error of the different algorithms {applied to the ICMEs in the test data set}. Data from the sheath region and the first 4~hours of the magnetic obstacle were used to train these models.}
\label{fig:scatterbz}
\end{figure}

\begin{figure}
\includegraphics[width=0.99\textwidth]{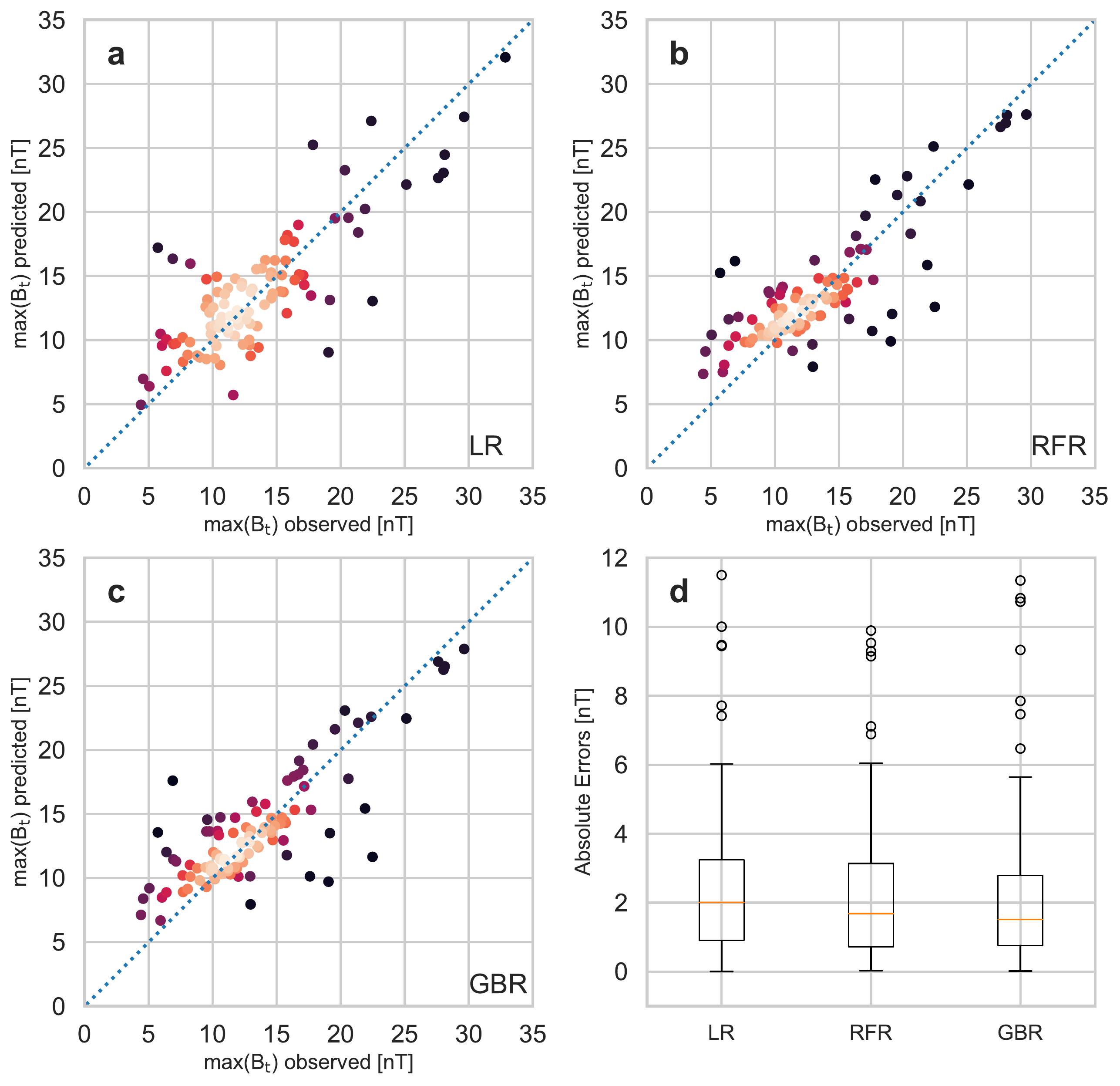}
\caption{Comparison of the predicted and observed maximum value of the total magnetic field with three different machine learning algorithms, where the color scales with the density of the data points. ({a}) Linear Regressor; ({b}) Random Forest Regressor; ({c}) Gradient Boosting Regressor; ({d}) absolute error of the different algorithms {applied to the ICMEs in the test data set}. Data from the sheath region and the first 4~hours of the magnetic obstacle were used to train these models.}
\label{fig:scatterbt}
\end{figure}

To illustrate the skill of the predictive tool, we plot the prediction of the minimum value of the B$_\textrm{z}$ component versus the observation. For this illustration, we feed data from the sheath region and 4~hours from the beginning of the magnetic obstacle into the {three trained} machine learning algorithms. {Figure~\ref{fig:scatterbz}(a--c)} shows a comparison of the prediction and observation of min(B$_\textrm{z}$) for the linear regressor (LR), random forest regressor (RFR), and the gradient boosting regressor (GBR), respectively. {Figure~\ref{fig:scatterbz}(d)} shows the boxplot for the absolute error of the different algorithms. Here the red line is the median error (50\textsuperscript{th} percentile), the edges are the 25\textsuperscript{th} and 75\textsuperscript{th} percentiles, the whiskers show the $\pm 2.5 \sigma$ range covering 99.3\% of the data, and the outliers are plotted as black circles. Agreement of the data points with the blue dashed lines in {Figure~\ref{fig:scatterbz}(a--c)} indicate the best performance. In comparison, the PCC for the LR, RFR, and GBR is $0.32$, $0.70$, and $0.71$, respectively. {In addition to the PCCs, we compute the p-values to test the null hypothesis that the relation between prediction and measurement is statistically insignificant. While a low p-value ($<0.05$) indicates that we can reject this null hypothesis, a larger p-value confirms it. We find that all the predictions are statistically significant because the p-value for the LR is equal to 0.001, and the p-values for the RFR and GBR are equal to 0. Although the LR results do not look promising, more reasonable results are achieved by the RFR and GBR.}

{Figure~\ref{fig:scatterbt}} shows the same illustration for the maximum value of the total magnetic field max(B$_\textrm{t}$). All three algorithms show considerably higher accuracy in comparison to the previous results for min(B$_\textrm{z}$). In particular, the PCC for the LR, RFR, and GBR is $0.54$, $0.88$, and $0.91$, respectively. {Again, we find that the relation between observation and prediction is statistically significant for all the algorithms because the p-values are equal to 0.} Both the RFR and GBR show promise for predicting max(B$_\textrm{t}$) as an estimate of the B$_\textrm{z}$ component. 

{For the 105~ICMEs under scrutiny in this section, we found that the predictive skill of the machine learning algorithms stays nearly constant throughout the solar cycle, and there are no significant differences among the three spacecraft.}

\begin{table*}[t]
\caption{Predictive abilities in terms of arithmetic mean (AM), standard deviation (SD), mean error (ME), mean absolute error (MAE), root mean square error (RMSE), the skill score (SS) relative to the simple mean value of all events, and the Pearson correlation coefficient (PCC). We use the sheath plus the first 4~hours of the MO to predict min(B$_\textrm{z}$) and max(B$_\textrm{t}$), respectively.}
\begin{center}
\begin{tabular}{llcccccccccc}
Target & Model                   & {AM}      & SD        & ME        & MAE       & RMSE      & SS        & PCC\\ 
         & & {[}\si{nT}{]} & {[}\si{nT}{]} & {[}\si{nT}{]} & {[}\si{nT}{]} & {[}\si{nT}{]} & \\\hline 
min(B$_\textrm{z}$) & LR & -9.67 & 12.26 &0.61 & 4.80 & 11.97 & -2.24 & 0.32 \\
min(B$_\textrm{z}$) & RFR & -9.47 & 4.58 &0.41 & 3.16 & 4.73 & 0.49 & 0.70 \\
min(B$_\textrm{z}$) & GBR & -9.58 & 4.12 &0.52 & 3.12 & 4.77 & 0.49 & 0.71 \\
min(B$_\textrm{z}$) & Observation 	& -9.06     & 6.65 & - & - & - & - &- \\ \hline
max(B$_\textrm{t}$) & LR & 14.18 & 11.15 &-0.28 & 3.64 & 9.48 & -0.55 & 0.54 \\
max(B$_\textrm{t}$) & RFR & 14.09 & 5.68 &-0.19 & 2.41 & 3.79 & 0.75 & 0.88 \\
max(B$_\textrm{t}$) & GBR & 14.18 & 6.33 &-0.29 & 2.23 & 3.20 & 0.82 & 0.91 \\
max(B$_\textrm{t}$) & Observation 	& 13.89     & 7.63 & - & - & - & - &- \\
\end{tabular}
\end{center}
\label{tab:p2p-results}
\end{table*}

For a more detailed error analysis, we compute point-to-point metrics as defined in {Table~\ref{tab:errorfunctions}}. The first three rows in {Table~\ref{tab:p2p-results}} show the results of the different algorithms for min(\Bz{}), and the columns show the corresponding values. The {fourth row} in the table shows the arithmetic mean (AM) and the standard deviation (SD) of the observation. From the results for min(\Bz{}), we see that the RFR and GBR outperform the LR in all metrics. While the MAE for the RFR and GBR is $3.16$~nT and $3.12$~nT, the MAE for the LR is $4.80$~nT. Also, the PCC for the RFR and GBR is $0.70$ and $0.71$, the PCC for the LR is $0.32$. This trend is confirmed in the SS comparing the skill to a baseline model. We find that the LR is worse than the baseline model with negative SS, and the results for the RFR and GBR are $0.49$, where $0$ would indicate the same skill as the baseline, and $1$ would indicate an optimum forecast. 

The last four rows in {Table~\ref{tab:p2p-results}} show the results of the different machine learning models for predicting max(B$_\textrm{t}$). Again we find that the RFR and GBR outperform the LR. For example, the MAE for the RFR and GBR is $2.41$~nT and $2.23$~nT, the MAE for the LR is $3.64$~nT. Focusing on the differences between the RFR and GBR for both targets, we find that both perform similarly, with the GBR providing slightly better results.

\begin{figure}
\includegraphics[width=0.99\textwidth]{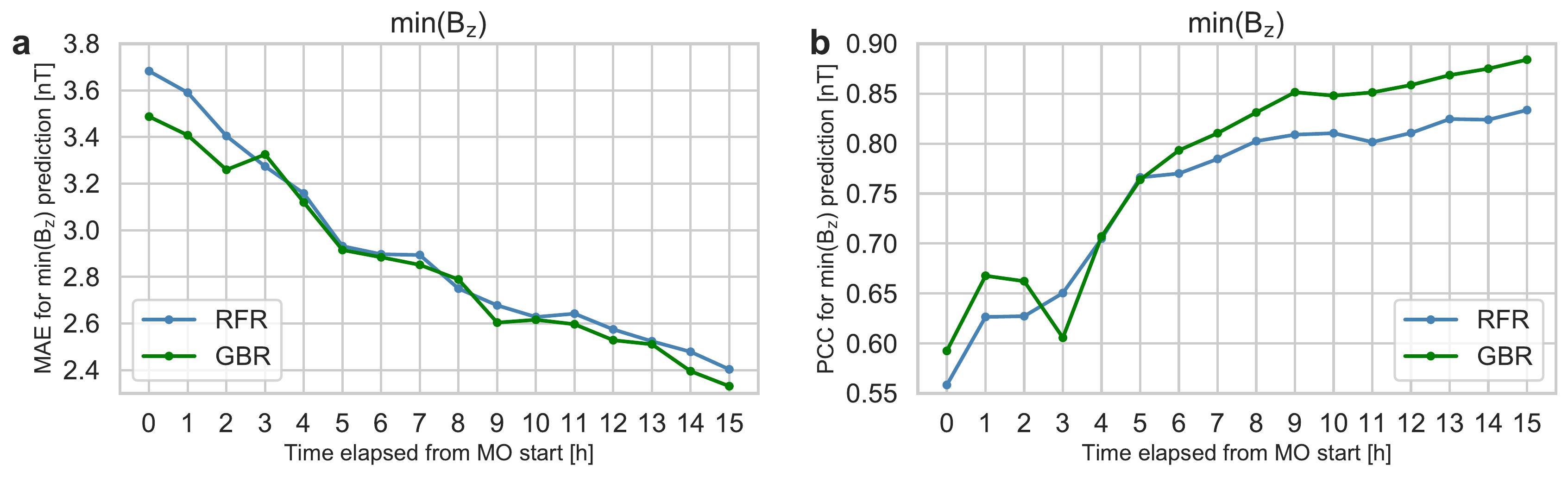}
\includegraphics[width=0.99\textwidth]{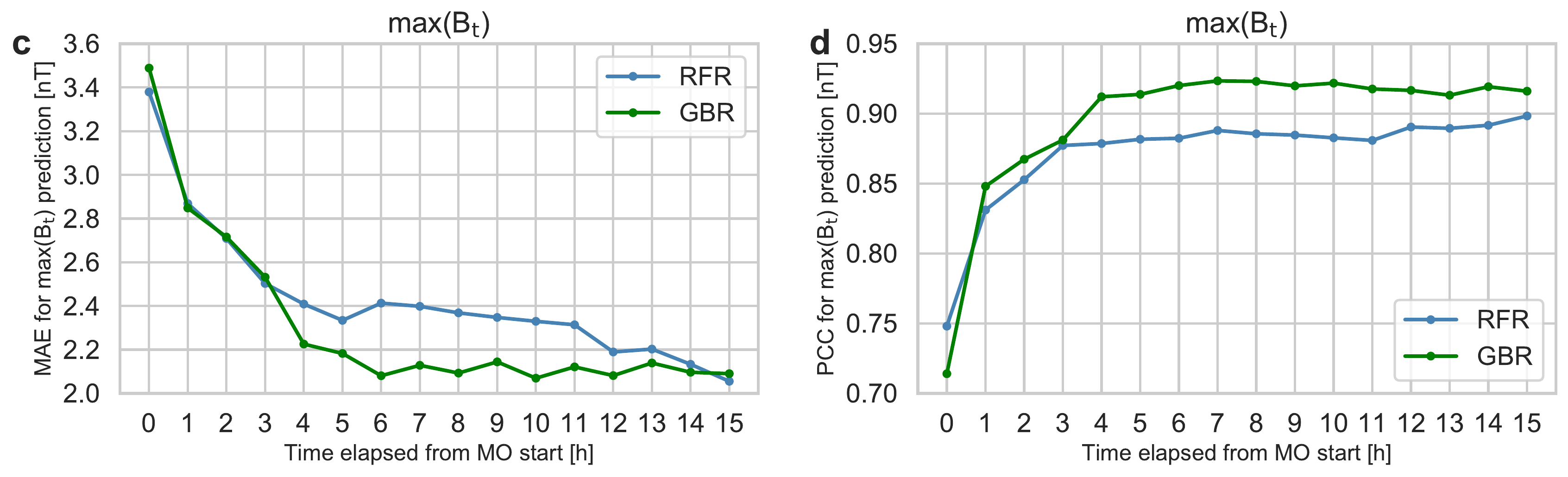}
\caption{Effect of the time elapsed from the magnetic obstacle start on the predictive ability of the different machine learning models. ({a}) MAE for the min(B$_\textrm{z}$) prediction; ({b}) PCC for the min(B$_\textrm{z}$) prediction; ({c}) MAE for the max(B$_\textrm{t}$) prediction; ({d}) PCC for the max(B$_\textrm{t}$) prediction.}
\label{fig:timeWindow}
\end{figure}

We furthermore investigate an operational setting where an ICME sweeps over the spacecraft. In this experimental real-time mode, we quantify how additional information from the magnetic flux rope improves the skill of the machine learning algorithms. {Figure~\ref{fig:timeWindow}} shows the skill of a string of trained RFR and GBR models for the prediction of min(B$_\textrm{z}$) and max(B$_\textrm{t}$). The first data point at 0~hours denotes the start of the magnetic obstacle where features are computed only from the sheath region. With increasing time elapsed from the magnetic obstacle start, more information from the target is included in the input data. As expected, we find that including more information from the magnetic obstacle improves the accuracy of the predictive skill. As an example, the PCC for the min(B$_\textrm{z}$) RFR prediction and max(B$_\textrm{t}$) RFR prediction improves in the first 4~hours by approximately 14\% and 13\%. We also find that more data from the magnetic obstacle improves the LR predictions (not shown in the plot). However, the variations in the LR predictions are significantly larger in comparison to the RFR and GBR when seeing more from the magnetic obstacle.

We complement this analysis with an event-based validation study. {We specifically use the mean value of the test data with 105~ICMEs as an event threshold.} Values below a threshold of $-9.06$~nT for min(B$_\textrm{z}$) and above $13.89$~nT for max(B$_\textrm{t}$) signify an event, and all other values are not events. As discussed in {Section~3.2}, the results of this analysis can be summarized in a contingency table, from which we can compute the measures defined in {Table~2}. {Table~4} lists the resulting measures for the different learners. In terms of an actionable prediction, the RFR provides the best results with a TSS of $0.30$, followed by the GBR with a TSS of $0.28$, and the LR with a TSS of $0.18$. Again we find a clear gap between the LR and the other two models. These differences indicate that the results by the LR are not only less promising in predicting the exact values but also in predicting the severity of the ICME.

In quantitative measures, we find that the GBR was able to predict if min(B$_\textrm{z}$) is below $-9.06$~nT in 77\% of all observed ICMEs in this category. On the other hand, 47\% of all the events above $-9.06$~nT were erroneously classified as being below this threshold by the GBR. When focusing on max(B$_\textrm{t}$) as an approximation, the results are significantly better. More specifically, the GBR was able to predict if max(B$_\textrm{t}$) is above $13.89$~nT in 77\% of all observed ICMEs in this category, and only 11\% of all the events below $13.89$~nT were erroneously classified.

{For end-users interested only in the most extreme events, we have conducted an additional analysis where we focus only on the 15 most decisive events in terms of min(B$_\textrm{z}$) and max(B$_\textrm{t}$). For the 15 strongest observed events in terms of min(B$_\textrm{z}$) below a TH of -13.93~nT, we find that the LR, RFR, and GBR can correctly predict 8, 9, and 10 events below this event threshold, respectively. Focusing on the 15 strongest events in max(B$_\textrm{t}$) above a TH of 19.05~nT, we find that the LR, RFR, and GBR can correctly predict 12, 11, and 11 events, respectively.}

{In summary, we find that the RFR and GBR provide promising first results and justify their use. While not solving the B$_\textrm{z}$ problem, this prototype predictive tool} can narrow down the extent of the B$_\textrm{z}$ component expected during the ICME arrival at Earth. We find that the predictive ability of the tool increases considerably with additional information as an ICME sweeps over the spacecraft.

\begin{table*}[t]
\caption{Contingency table and skill measures for an event-based validation analysis. The table shows the number of observed (N$_\textrm{obs}$) and predicted (N$_\textrm{pred}$) events, hits ({true positives}; TPs), false alarms ({false positives}; FPs), misses ({false negatives}, FNs), correct rejections ({true negatives}, TNs), and the metrics derived from these entries including: the true positive rate (TPR) and false positive rate (FPR), threat score (TS), true skill statistics (TSS), and bias (BS).}
\begin{center}
\begin{tabular}{llccccccccccc}
Target & Model & N$_\textrm{obs}$ & N$_\textrm{pred}$ & TP & FP & FN & TN  & TPR & FPR & TS & TSS & BS \\\hline
min(B$_\textrm{z}$) & LR & 60 & 60 & 39 & 21 & 21 & 24 & 0.65 & 0.47 & 0.48 & 0.18 & 1.00 \\
min(B$_\textrm{z}$) & RFR & 60 & 67 & 46 & 21 & 14 & 24 & 0.77 & 0.47 & 0.57 & 0.30 & 1.12 \\
min(B$_\textrm{z}$) & GBR & 60 & 68 & 46 & 22 & 14 & 23 & 0.77 & 0.49 & 0.56 & 0.28 & 1.13 \\\hline
max(B$_\textrm{t}$) & LR & 39 & 41 & 28 & 13 & 11 & 53 & 0.72 & 0.20 & 0.54 & 0.52 & 1.05 \\
max(B$_\textrm{t}$) & RFR & 39 & 34 & 28 & 6 & 11 & 60 & 0.72 & 0.09 & 0.62 & 0.63 & 0.87 \\
max(B$_\textrm{t}$) & GBR & 39 & 37 & 30 & 7 & 9 & 59 & 0.77 & 0.11 & 0.65 & 0.66 & 0.95 \\
\end{tabular}
\end{center}
\label{tab:2}
\end{table*}

% ----------------------------------------------------------------------
\section{Discussion}\label{sec:discussion}
% ----------------------------------------------------------------------
A traditional way of approaching the B$_\textrm{z}$ problem is to simulate the propagation of ICMEs with MHD codes. Examples are SUSANOO-CME~\cite{shiota2016}, AWSoM-SWMF~\cite{jin2017}, MAS~\cite{torok2018}, and EUHFORIA \cite{poedts_2020}. These MHD codes include the magnetic flux rope structure and simulate the ICME evolution from near the Sun into interplanetary space. Although a full physics-based description is desirable, a breakthrough in solving the B$_\textrm{z}$ problem with MHD codes alone is challenging. On the one hand, the boundary conditions for all MHD codes are based on solar magnetic field measurements, but these are known to have large inherent uncertainties. {Additionally, small uncertainties in the initial conditions of a CME grow to large errors in the predicted magnetic field components at 1~AU, demonstrated e.g. with semi-empirical forward models by \citeA{kay2018} and \citeA{Moestl_2018}}. Even with flux-transport models based on Sun-Earth line observations, it is difficult to develop a time-dependent coronal magnetic model, which is needed to capture the early CME evolution. On the other hand, the solar corona is not strictly an MHD environment. While an MHD simulation can describe the large-scale coronal conditions within the Alfvén surface, the evolution of CMEs occurs over a broad range of spatial scales. Small-scale processes such as triggering instabilities are essential to capture the whole picture. Without a definitive physical solution, we need to study new predictive tools that can enhance B$_\textrm{z}$ prediction capabilities, and ideally, provide constraints for physics-based models. 

In this context, we have studied the hypothesis that upstream in situ measurements of the sheath region and the first few hours of the magnetic flux rope of the ICME are useful to predict estimates of the B$_\textrm{z}$ component. To test this hypothesis, we developed a predictive tool based on machine learning that is trained and tested on 348 ICMEs. We found that the tool can predict the minimum value of the B$_\textrm{z}$ component (PCC$=0.71$) and the maximum value of the total magnetic field B$_\textrm{t}$ (PCC$=0.91$) in the magnetic obstacle in reasonable agreement with observations. While the investigated hypothesis certainly does not solve the B$_\textrm{z}$ problem, it shows promising first results for ICMEs that have a magnetic flux rope signature, and its application might be useful for operational space weather forecasting. {For an application in an operational context, we computed the best-case warning time when we assume immediate data access and model application to a real ICME. To do so, we compute the lead time between the arrival of the magnetic obstacle and the peak in the min(B$_\textrm{z}$) and max(B$_\textrm{t}$) values. For min(B$_\textrm{z}$), we find that the average warning time is 9.68~h, the median warning time is 5.87~h, and the maximum warning time is 46.15~h. The same analysis for max(B$_\textrm{t}$) showed that the average warning is 5.98~h, the median warning time is 2.53~h, and the maximum warning time is 33.55~h.}

To put our results into context, we want to discuss the following three limitations of our study. First, we have significantly reduced the complexity of the B$_\textrm{z}$ problem by focusing on estimates of the B$_\textrm{z}$ component for the whole magnetic flux rope, specifically min(B$_\textrm{z}$) and max(\Bt{}). Focusing on these large-scale statistics of the magnetic flux rope is considerably easier than predicting the temporal evolution of B$_\textrm{z}$. Nevertheless, the ability of our prototype to predict min(B$_\textrm{z}$), a proxy for ICME geoeffectiveness, is essential in an actionable forecast. 

Second, we have trained and tested the machine learning algorithms on ICMEs that occurred between the years 2007 to 2021 using the ICMECATv2.0 catalog. The main criterion was that the ICMEs needed to show either sheath region signatures or a density pileup in front of a magnetic obstacle. This criterion reduced the number of ICMEs to approximately $65\%$ of the original catalog. The expected magnetic flux rope signatures in this study include increasing magnetic field strength, rotation of at least one field component, and plasma-$\beta < 1$~\cite{burlaga1988}. These signatures are not always observed by satellites~\cite{rouillard2009, wood2009, vourlidas2014}. We know that flux ropes undergo deformation and distortion during their dynamic evolution, often leading to significant deviation from an idealized flux rope structure~\cite{riley2006, savani2010, nieves-chinchilla2012}. Therefore, an open question is how well we can extrapolate the predictive skill presented here to an operational space weather forecast.

Third and in the context of the last point, we know that the success of the predictive tool in any operational setting will depend on accurate automated ICME detection in mission data as introduced in~\citeA{telloni2019, nguyen19}. {Combining our predictive tool with an automated ICME detection algorithm introduces new challenges. For example, we would need to understand the effect of a timing error in the automated ICME detection on the coupled predictive tool for the B$_\textrm{z}$ prediction. More research is required to assess these uncertainties and understand their effect on the robustness and reliability of the B$_\textrm{z}$ prediction.}

In the future, we will work on several topics to improve upon this prototype. {Besides coupling our predictive tool with an automated ICME detection algorithm,} we will also work on new strategies to forecast the temporal evolution of the B$_\textrm{z}$ component. Here, our approach is two-fold. Initially, a data driven machine learning approach to predict the temporal evolution of the B$_\textrm{z}$ component is envisaged. Next, we will explore the integration of a semi-empirical magnetic flux rope model as discussed in~\citeA{weiss2021a} and \citeA{weiss2021b} in our framework. The advantage of this technique is that it can model the CME flux ropes in conjunction with an approximate Bayesian computation (ABC) algorithm that fits the model to in situ magnetic field measurements. {We will study if the CME flux rope model can fit} the rest of the magnetic flux rope from the first few hours of the ICME. For both approaches, we aim to deduce error boundaries for the B$_\textrm{z}$ prediction. 

In a long-term vision, our predictive tool could make use of mission data from a space weather monitor closer to the Sun using solar sail technology~\cite{west2004}. When the space weather monitor is placed near the Sun-Earth line at approximately $3 \times 10^6$~km upstream from Earth or twice the distance to L1, the warning time compared to L1 satellites is doubled. In other words, for an ICME traveling with 400~km/s (which is close to the median speed of the ICMEs under scrutiny), the lead time before the event arrives at Earth increases from approximately 1~hour to 2~hours. {Since the present methodology is developed and tested for three different spacecraft, our predictive tool could be expanded to future space missions orbiting at 1 AU when differences among instrumentation are taken into consideration.}

To allow the community to compare future studies with our findings, the source code, ICMECATv2.0 catalog, and related data are available online as outlined in Section~\ref{sec:sources}.

% ----------------------------------------------------------------------
\section{Summary}\label{sec:summary}
% ----------------------------------------------------------------------
The capacity of ICMEs to cause extreme geomagnetic storms fundamentally depends on their internal plasma structure and their B$_\textrm{z}$ magnetic field. At present, we can not predict the B$_\textrm{z}$ magnetic field component with sufficient warning time before the ICME arrival at Earth. This conundrum is often called the B$_\textrm{z}$ problem. 

We shine new light on the B$_\textrm{z}$ problem by studying the research question if upstream in situ measurements of the ICME sheath region and the first few hours of the magnetic flux rope are sufficient for predicting the B$_\textrm{z}$ component. To do so, we developed a predictive tool based on machine learning that is trained and tested on 348~ICME events observed by the Wind, STEREO-A, and STEREO-B spacecraft. We train machine learning models to output the minimum value of the B$_\textrm{z}$ component and the maximum value of the total magnetic field B$_\textrm{t}$ in the magnetic obstacle. 

To test our predictive tool in an experimental real-time mode, we let the ICMEs sweep over the spacecraft and assess how continually feeding new information into the tool improves the B$_\textrm{z}$ predictions. 

{Our study shows} that the predictive tool can predict the minimum value of the B$_\textrm{z}$ component (MAE$=3.12$~nT, PCC$=0.71$) and the maximum value of the total magnetic field B$_\textrm{t}$ (MAE$=2.23$~nT, PCC$=0.91$) in the magnetic obstacle in reasonable agreement with observations. While the investigated hypothesis does not solve the B$_\textrm{z}$ problem, the first version of the predictive tool shows reasonable results for ICMEs that have a clear magnetic flux rope signature, and its application might be suited for operational space weather forecasting in the future.

% ----------------------------------------------------------------------
\section{Data Availability Statement}\label{sec:sources}
% ----------------------------------------------------------------------
\noindent The solar wind in situ data are available as python numpy arrays at \url{https://doi.org/10.6084/m9.figshare.12058065.v8} (updated on 2021 April 29) and were originally downloaded from \url{https://stereo-ssc.nascom.nasa.gov} (STEREO) and \url{https://spdf.gsfc.nasa.gov/pub/data/wind/} (Wind). The current version of the ICME catalog ICMECATv2.0, version 6 updated on 2021 April 29 and as published on the data sharing platform figshare, was used in this study:  \url{https://doi.org/10.6084/m9.figshare.6356420.v6}. (The most up-to-date version can be found at \url{https://helioforecast.space/icmecat}.) The paper source code is available at \url{https://github.com/helioforecast/Papers/tree/master/Reiss2021_MLrope}.

\acknowledgments
M.A.R., C.M., R.L.B., U.V.A., T.A., A.J.W., and J.H.~thank the Austrian Science Fund (FWF): P31659-N27, P31521-N27, and P31265-N27. Europlanet 2024 RI has received funding from the European Union’s Horizon 2020 research and innovation programme under grant agreement No 871149.

\bibliography{reiss21b}

\newpage
\appendix

\end{document}